# Automatic Detection of Coronavirus Disease (COVID-19) Using X-ray Images and Deep Convolutional Neural Networks

Ali Narin[1], Ceren Kaya[2, *], Ziynet Pamuk[2]

[1] Department of Electrical and Electronics Engineering, Zonguldak Bulent Ecevit University, 67100, Zonguldak, Turkey. alinarin@beun.edu.tr

[2] Department of Biomedical Engineering, Zonguldak Bulent Ecevit University, 67100, Zonguldak, Turkey. ceren.kaya@beun.edu.tr, ziynet.pamuk@beun.edu.tr

*Corresponding author: ceren.kaya@beun.edu.tr, crnkaya@hotmail.com

**Abstract**

The 2019 novel coronavirus disease (COVID-19), with a starting point in China, has spread rapidly among people living in other countries, and is approaching approximately 34,986,502 cases worldwide according to the statistics of European Centre for Disease Prevention and Control. There are a limited number of COVID-19 test kits available in hospitals due to the increasing cases daily. Therefore, it is necessary to implement an automatic detection system as a quick alternative diagnosis option to prevent COVID-19 spreading among people. In this study, five pre-trained convolutional neural network based models (ResNet50, ResNet101, ResNet152, InceptionV3 and Inception-ResNetV2) have been proposed for the detection of coronavirus pneumonia infected patient using chest X-ray radiographs. We have implemented three different binary classifications with four classes (COVID-19, normal (healthy), viral pneumonia and bacterial pneumonia) by using 5-fold cross validation. Considering the performance results obtained, it has seen that the pre-trained ResNet50 model provides the highest classification performance (96.1% accuracy for Dataset-1, 99.5% accuracy for Dataset-2 and 99.7% accuracy for Dataset-3) among other four used models.

# 1. Introduction

The coronavirus disease (COVID-19) pandemic emerged in Wuhan, China in December 2019 and became a serious public health problem worldwide [1,2]. Until now, no specific drug or vaccine has been found against COVID-19 [2]. The virus that causes COVID-19 epidemic disease is called severe acute respiratory syndrome coronavirus-2 (SARS-CoV-2) [3]. Coronaviruses (CoV) is a large family of viruses that cause diseases such as Middle East Respiratory Syndrome (MERS-CoV) and Severe Acute Respiratory Syndrome (SARS-CoV). COVID-19 is a new species discovered in 2019 and has not been previously identified in humans [4]. COVID-19 causes lighter symptoms in about 99% of cases, according to early data, while the rest is severe or critical [5]. As of 4th October 2020, the total number of worldwide cases of Coronavirus is 35,248,330. Of these, 1,039,541 (4%) people were deaths and 26,225,235 (96%) were recovered. The number of active patients is 7,983,554. Of these, 7,917,287 (99%) had mild disease while 66,267 (1%) had more severe disease [6]. Nowadays the world is struggling with the COVID-19 epidemic. Deaths from pneumonia developing due to the SARS-CoV-2 virus are increasing day by day.

Chest radiography (X-ray) is one of the most important methods used for the diagnosis of pneumonia worldwide [7]. Chest X-ray is a fast, cheap [8] and common clinical method [9-11]. The chest X-ray gives the patient a lower radiation dose compared to computed tomography (CT) and magnetic resonance imaging (MRI) [11]. However, making the correct diagnosis from X-ray images requires expert knowledge and experience [7]. It is much more difficult to diagnose using a chest X-ray than other imaging modalities such as CT or MRI [8].

By looking at the chest X-ray, COVID-19 can only be diagnosed by specialist physicians. The number of specialists who can make this diagnosis is less than the number of normal doctors. Even in normal times, the number of doctors per person is insufficient in countries around the world. According to data from 2017, Greece ranks first with 607 doctors per 100,000 people. In other countries, this number is much lower [12].

In case of disasters such as COVID-19 pandemic, demanding health services at the same time, collapse of the health system is inevitable due to the insufficient number of hospital beds and health personnel. Also, COVID-19 is a highly contagious disease, and doctors, nurses, and caregivers are most at risk. Early diagnosis of pneumonia has a vital importance both in terms of slowing the speed of the spread of the epidemic by quarantining the patient and in the recovery process of the patient [13].

Doctors can diagnose pneumonia from the chest X-ray more quickly and accurately thanks to computer-aided diagnosis (CAD) [8]. Use of artificial intelligence methods are increasing due to its ability to cope with enormous datasets exceeding human potential in the field of medical services [14]. Integrating CAD methods into radiologist diagnostic systems greatly reduces the workload of doctors and increases reliability and quantitative analysis [11]. CAD systems based on deep learning and medical imaging are becoming more and more research fields [14,15].

## 2. Related Works

Studies diagnosed with COVID-19 using chest X-rays have binary or multiple classifications. Some studies use raw data while others have feature extraction process. The number of data used in studies also varies. Among the studies, the most preferred method is convolutional neural network (CNN).

Apostolopoulos and Bessiana used a common pneumonia, COVID-19-induced pneumonia, and an evolutionary neural network for healthy differentiation on automatic detection of COVID-19. In particular, the procedure called transfer learning has been adopted. With transfer learning, the detection of various abnormalities in small medical image datasets is an achievable goal, often with remarkable results [16]. Based on chest X-ray images, Zhang et al. aimed to develop a deep learning-based model that can detect COVID-19 with high sensitivity, providing fast and reliable scanning [17]. Singh et al. classified the chest computed tomography (CT) images from infected people with and without COVID-19 using multi-objective differential evolution (MODE) based CNN [18]. In the study of Chen et al, they proposed Residual Attention U-Net for automated multi class segmentation technique to prepare the ground for the quantitative diagnosis of lung infection on COVID-19 related pneumonia using CT images [19]. Adhikari's study suggested a network called "Auto Diagnostic Medical Analysis" trying to find infectious areas to help the doctor better identify the diseased part, if any. Both X-ray and CT images were used in the study. It has been recommended DenseNet network to remove and mark infected areas of the lung [20]. In the study by Alqudah et al., two different methods were used to diagnose COVID-19 using chest X-ray images. The first one used AOCTNet, MobileNet and ShuffleNet CNNs. Secondly, the features of their images have been removed and they have been classified using softmax classifier, K nearest neighbor (kNN), Support vector machine (SVM) and Random forest (RF) algorithms [21]. Khan et al. classified the chest X-ray images from normal, bacterial and viral pneumonia cases using the Xception architecture to detect COVID-19 infection [22]. Ghoshal and Tucker used the dropweights based Bayesian CNN model using chest X-ray images for the diagnosis of COVID-19 [23]. Hemdan et al. used VGG19 and DenseNet models to diagnose COVID-19 from X-ray images [24]. Uçar and Korkmaz worked on X-ray images for COVID-19 diagnosis and supported the SqueezeNet model with Bayesian optimization [25]. In the study conducted by Apostopolus et al., they performed automatic detection from X-ray images using CNNs with transfer learning [26]. Sahinbas and Catak used X-ray images for the diagnosis of COVID-19 and worked on VGG16, VGG19, ResNet, DenseNet and InceptionV3 models [27]. Medhi et al. used X-ray images as feature extraction and segmentation in their study, then COVID-19 was positively and normally classified using CNN [28]. Barstugan et al. classified X-ray images for the diagnosis of COVID-19 using five different feature extraction methods that are Grey Level Cooccurrence Matrix (GLCM), Local Directional Patterns (LDP), Grey Level Run Length Matrix (GLRLM), Grey Level Size Zone Matrix (GLSZM), and Discrete Wavelet Transform (DWT). The obtained features were classified by SVM. During the classification process, 2-fold, 5-fold, and 10-fold cross-validation methods were used [29]. Punn and Agarwal worked on X-ray images and used ResNet, InceptionV3, InceptionResNet models to diagnose COVID-19 [30]. Afshar et al. developed deep neural network (DNN) based diagnostic solutions and offered an alternative modeling framework based on Capsule Networks that can process on small data sets [31].

In our previous study in March 2020, we used ResNet50, InceptionV3 and Inception-ResNetV2 models for the diagnosis of COVID-19 using chest X-ray images. However, since there was not enough data on COVID-19, we were only able to train through 50 normal and 50 COVID-19 positive cases [32].

In this study, we have proposed an automatic prediction of COVID-19 using a deep convolution neural network based pre-trained transfer models and chest X-ray images. For this purpose, we have used ResNet50, ResNet101, ResNet152, InceptionV3 and Inception-ResNetV2 pre-trained models to obtain higher prediction accuracies for three different binary datasets including X-ray images of normal (healthy), COVID-19, bacterial and viral pneumonia patients.

The novelty and originality of proposed study is summarized as follows:

**i)** The proposed models have end-to-end structure without manual feature extraction, selection and classification.

**ii)** The performances of the COVID-19 data across normal, viral pneumonia and bacterial pneumonia classes were significantly higher.

**iii)** It has been studied with more data than many studies in the literature.

**iv)** It has been studied and compared with 5 different CNN models.

**v)** A high-accuracy decision support system has been proposed to radiologists for the automatic diagnosis and detection of patients with suspected COVID-19 and follow-up.

The flow of the manuscript is organized as follows: Dataset is expressed in detail in Section 3.1. Deep transfer learning architecture, pre-trained models and experimental setup parameters are described in Section 3.2 and 3.3, respectively. Performance metrics are given in detail in Section 3.4. Obtained experimental results from proposed models and discussion are presented in Section 4 and 5, respectively. Finally, in Section 6, the conclusion and future works are summarized.

## 3. Materials and Methods

### 3.1 Dataset

In this study, chest X-ray images of 341 COVID-19 patients have been obtained from the open source GitHub repository shared by Dr. Joseph Cohen et al. [33]. This repository is consisting chest X-ray / computed tomography (CT) images of mainly patients with acute respiratory distress syndrome (ARDS), COVID-19, Middle East respiratory syndrome (MERS), pneumonia, severe acute respiratory syndrome (SARS). 2800 normal (healthy) chest X-ray images were selected from "ChestX-ray8" database [34]. In addition, 2772 bacterial and 1493 viral pneumonia chest X-ray images were used from Kaggle repository called "Chest X-Ray Images (Pneumonia)" [35].

Our experiments have been based on three binary created datasets (Dataset-1, Dataset-2 and Dataset-3) with chest X-ray images. Distribution of images per class in created datasets are given Table 1.

**Table 1.** Number of images per class for each dataset.

| Classes / Datasets | Bacterial Pneumonia | COVID-19 | Normal | Viral Pneumonia |
|---|---|---|---|---|
| Dataset-1 | - | 341 | 2800 | - |
| Dataset-2 | - | 341 | - | 1493 |
| Dataset-3 | 2772 | 341 | - | - |

All images were resized to 224x224 pixel size in the datasets. In Figure 1, representative chest X-ray images of normal (healthy), COVID-19, bacterial and viral pneumonia patients are given, respectively.

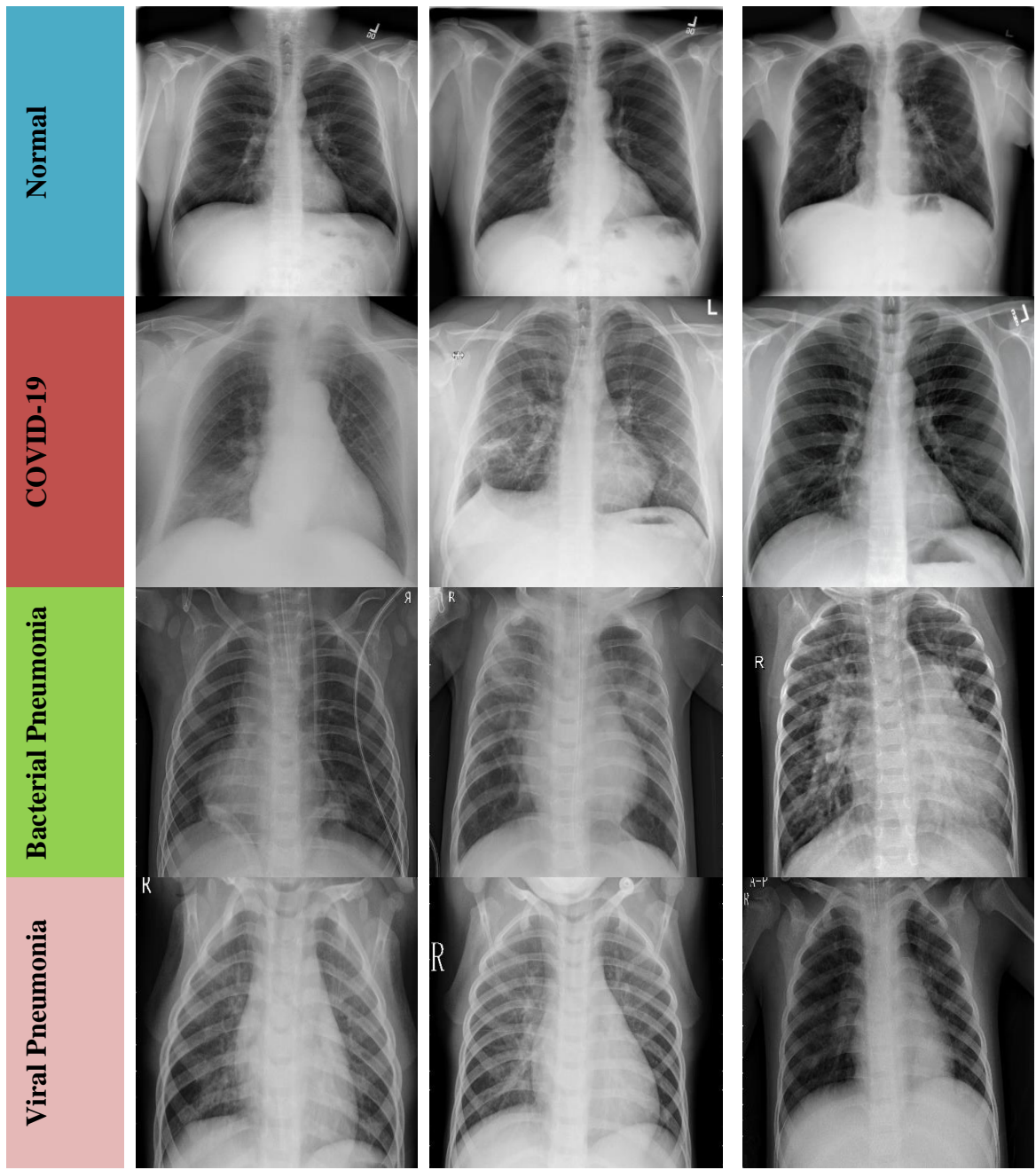

**Figure 1.** Representative chest X-ray images of normal (healthy) (first row), COVID-19 (second row), bacterial (third row) and viral pneumonia (fourth row) patients.

## 3.2 Architecture of Deep Transfer Learning

Deep learning is a sub-branch of the machine learning field, inspired by the structure of the brain. Deep learning techniques used in recent years continue to show an impressive performance in the field of medical image processing, as in many fields. By applying deep learning techniques to medical data, it is tried to draw meaningful results from medical data.

Deep learning models have been used successfully in many areas such as classification, segmentation and lesion detection of medical data. Analysis of image and signal data obtained with medical imaging techniques such as Magnetic Resonance Imaging (MRI), Computed Tomography (CT) and X-ray with the help of deep learning models. As a result of these analyzes, detection and diagnosis of diseases such as diabetes mellitus, brain tumor, skin cancer and breast cancer are provided with convenience [36-41].

A convolutional neural network (CNN) is a class of deep neural networks used in image recognition problems [42]. Coming to how CNN works, the images given as input must be recognized by computers and converted into a format that can be processed. For this reason, images are first converted to matrix format. The system determines which image belongs to which label based on the differences in images and therefore in matrices. It learns the effects of these differences on the label during the training phase and then makes predictions for new images using them. CNN consists of three different layers that are a convolutional layer, pooling layer, and fully connected layer to perform these operations effectively. The feature extraction process takes place in both convolutional and pooling layers. On the other hand, the classification process occurs in fully connected layer. These layers are examined sequentially in the following.

### 3.2.1 Convolutional Layer

Convolutional layer is the base layer of CNN. It is responsible for determining the features of the pattern. In this layer, the input image is passed through a filter. The values resulting from filtering consist of the feature map. This layer applies some kernels that slide through the pattern to extract low- and high-level features in the pattern [43]. The kernel is a 3x3 or 5x5 shaped matrix to be transformed with the input pattern matrix. Stride parameter is the number of steps tuned for shifting over input matrix. The output of convolutional layer can be given as:

$$x_j^l = f\left( \sum_{a=1}^{N} w_j^{l-1} * y_a^{l-1} + b_j^l \right) \qquad (1)$$

where $x_j^l$ is the j-th feature map in layer l, $w_j^{l-1}$ indicates j-th kernels in layer l-1, $y_a^{l-1}$ represents the a-th feature map in layer l-1, $b_j^l$ indicates the bias of the j-th feature map in layer l, N is number of total features in layer l-1, and (*) represents vector convolution process.

### 3.2.2 Pooling Layer

The second layer after the convolutional layer is the pooling layer. Pooling layer is usually applied to the created feature maps for reducing the number of feature maps and network parameters by applying corresponding mathematical computation. In this study, we used max-pooling and global average pooling. The max-pooling process selects only the maximum value by using the matrix size specified in each feature map, resulting in reduced output neurons. There is also a global average pooling layer that is only used before the fully connected layer, reducing data to a single dimension. It is connected to the fully connected layer after global average pooling layer. The other intermediate layer used is the dropout layer. The main purpose of this layer is to prevent network overfitting and divergence [44].

### 3.2.3 Fully Connected Layer

Fully connected layer is the last and most important layer of CNN. This layer functions like a multi-layer perceptron. Rectified Linear Unit (ReLU) activation function is commonly used on fully connected layer, while Softmax activation function is used to predict output images in the last layer of fully connected layer. Mathematical computation of these two activation functions are as follow:

$$ReLU(x) = \begin{cases} 0, & x < 0 \\ x, & x \geq 0 \end{cases} \tag{2}$$

$$Soft\max(x_i) = \frac{e^{x_i}}{\sum_{y=1}^{m} e^{x_y}} \tag{3}$$

where $x_i$ and m represent input data and the number of classes, respectively.

Neurons in a fully connected layer have full connections to all activation functions in previous layer.

### 3.2.4. Pre-Trained Models

In the analysis of medical data, one of the biggest difficulties faced by researchers is the limited number of available datasets. Deep learning models often need a lot of data. Labeling this data by experts is both costly and time consuming. The biggest advantage of using transfer learning method is that it allows the training of data with fewer datasets and requires less calculation costs. With the transfer learning method, which is widely used in the field of deep learning, the information gained by the pre-trained model on a large dataset is transferred to the model to be trained.

In this study, we built deep CNN based ResNet50, ResNet101, ResNet152, InceptionV3 and Inception-ResNetV2 models for the classification of COVID-19 Chest X-ray images to three different binary classes (Binary Class-1 = COVID-19 and normal (healthy), Binary Class-2 = COVID-19 and viral pneumonia, Binary Class-3 = COVID-19 and bacterial pneumonia). In addition, we applied transfer learning technique that was realized by using ImageNet data to overcome the insufficient data and training time. The schematic representation of conventional CNN including pre-trained ResNet50, ResNet101, ResNet152, InceptionV3 and Inception ResNetV2 models for the prediction of normal (healthy), COVID-19, bacterial and viral pneumonia patients were depicted in Figure 2. It is also available publicly for open access at https://github.com/drcerenkaya/COVID-19-DetectionV2.

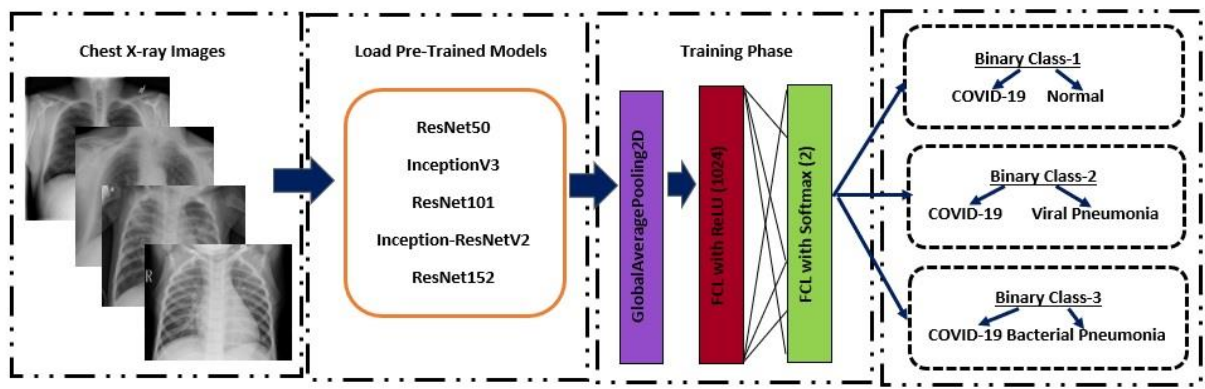

**Figure 2.** Schematic representation of pre-trained models for the prediction of normal (healthy), COVID-19, bacterial and viral pneumonia patients.

### ResNet50

Residual neural network (ResNet) model is an improved version of convolutional neural network (CNN). ResNet adds shortcuts between layers to solve a problem. Thanks to this, it prevents the distortion that occurs as the network gets deeper and more complex. In addition, bottleneck blocks are used to make training faster in the ResNet model [45]. ResNet50 is a 50-layer network trained on the ImageNet dataset. ImageNet is an image database with more than 14 million images belonging to more than 20 thousand categories created for image recognition competitions [46].

### InceptionV3

InceptionV3 is a kind of convolutional neural network model. It consists of numerous convolution and maximum pooling steps. In the last stage, it contains a fully connected neural network [47]. As with the ResNet50 model, the network is trained with ImageNet dataset.

**Inception-ResNetV2**

The model consists of a deep convolutional network using the Inception-ResNetV2 architecture that was trained on the ImageNet-2012 dataset. The input to the model is a 299×299 image, and the output is a list of estimated class probabilities [48].

**ResNet101 & ResNet152**

ResNet101 and ResNet152 consist of 101 and 152 layers respectively due to stacked ResNet building blocks. You can load a pretrained version of the network trained on more than a million images from the ImageNet database [46]. As a result, the network has learned rich feature representations for a wide range of images. The network has an image input size of 224x224.

## 3.3 Experimental Setup

Python programming language was used to train the proposed deep transfer learning models. All experiments were performed on Google Colaboratory (Colab) Linux server with the Ubuntu 16.04 operating system using the online cloud service with Central Processing Unit (CPU), Tesla K80 Graphics Processing Unit (GPU) or Tensor Processing Unit (TPU) hardware for free. CNN models (ResNet50, ResNet101, ResNet152, InceptionV3 and Inception-ResNetV2) were pre-trained with random initialization weights by optimizing the cross-entropy function with adaptive moment estimation (ADAM) optimizer ($\beta_1 = 0.9$ and $\beta_2 = 0.999$). The batch size, learning rate and number of epochs were experimentally set to 3, 1e-5 and 30, respectively for all experiments. The dataset used was randomly split into two independent datasets with 80% and 20% for training and testing respectively. As cross validation method, k-fold was chosen and results were obtained according to 5 different k values (k=1-5) as shown in Figure 3.

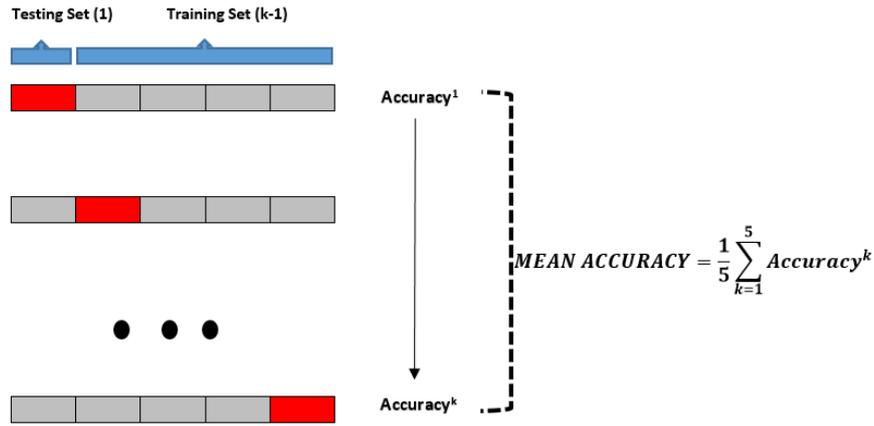

**Figure 3.** Visual display of testing and training datasets for 5-fold cross validation.

### 3.4 Performance Metrics

5 criteria were used for the performances of deep transfer learning models. These are:

$$\text{Accuracy} = (TN + TP) / (TN + TP + FN + FP) \tag{4}$$

$$\text{Recall} = TP / (TP + FN) \tag{5}$$

$$\text{Specificity} = TN / (TN + FP) \tag{6}$$

$$\text{Precision} = TP / (TP + FP) \tag{7}$$

$$\text{F1-Score} = 2 \times ((\text{Precision} \times \text{Recall})/(\text{Precision}+\text{Recall})) \tag{8}$$

TP, FP, TN and FN given in Equation (4) – (8) represent the number of True Positive, False Positive, True Negative and False Negative, respectively. For Dataset-1; given a test dataset and model, TP is the proportion of positive (COVID-19) that are correctly labeled as COVID-19 by the model; FP is the proportion of negative (normal) that are mislabeled as positive (COVID-19); TN is the proportion of negative (normal) that are correctly labeled as normal and FN is the proportion of positive (COVID-19) that are mislabeled as negative (normal) by the model.

### 4. Experimental Results

In this paper, we performed 3-different binary classifications with 4 different classes (COVID-19, normal, viral pneumonia and bacterial pneumonia). 5-fold cross validation method has been used in order to get a robust result in this study performed with 5-different pre-trained models that are InceptionV3, ResNet50, ResNet101, ResNet152 and Inception-ResNetV2. While 80% of the data is reserved for training, the remaining 20% is reserved for testing. All this process continued until each 20% part was tested.

Firstly, the accuracy and loss values in the training process obtained for the models applied to Dataset-1 that includes Binary Class-1 (COVID-19 / Normal classes) are given in Figure 4 and Figure 5. It is clear that the performance of the ResNet50 model is better than the other models. It can be said that the ResNet50 model reaches lower values among the loss values of other models. Detection performance on test data is shown in Figure 6. While a lot of oscillation is observed in some models, some models are more stable. The ResNet50 model appears to have less oscillation after the 15$^{th}$ epoch. Comprehensive performance values for each fold value of each model are given in Table 2. As seen from Table 2 that the detection of the ResNet50 model in the COVID-19 class is significantly higher than the other models. ResNet50 and ResNet101 have the highest overall performance with 96.1%. It is obvious that the excess of normal data results in higher performance in all models.

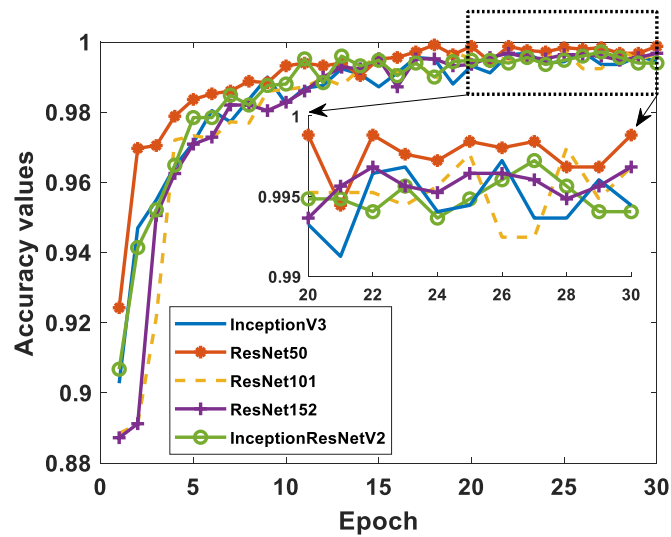

**Figure 4.** Binary Class-1: Comparison of training accuracy of 5 different models for fold-4.

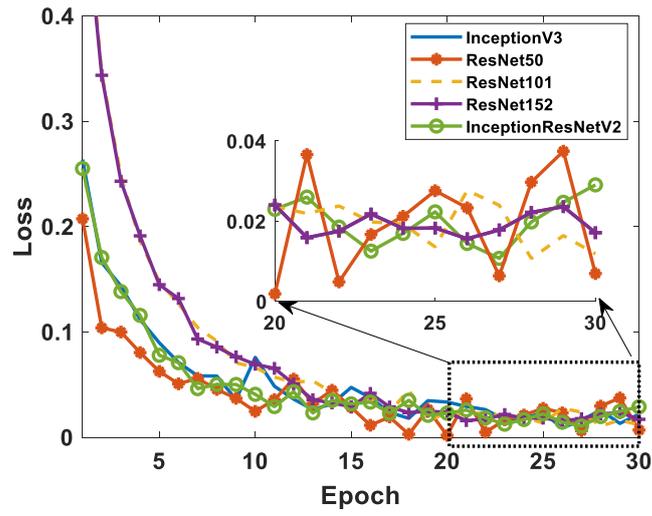

**Figure 5.** Binary Class-1: Comparison of training loss values of 5 different models for fold-4.

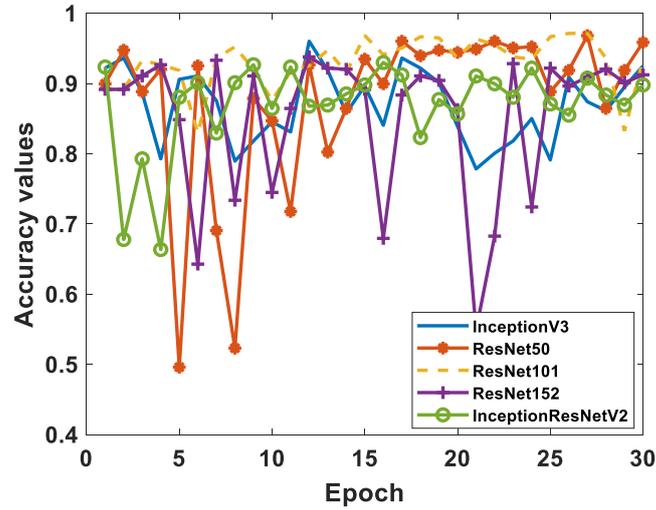

**Figure 6.** Binary Class-1: Comparison of testing accuracy of 5 different models for fold-4.

**Table 2.** All performances of 5 different models on each fold for COVID-19 / Normal binary classification. The abbreviations in Table 2 are: True Positive (TP), True Negative (TN), False Positive (FP), False Negative (FN), Accuracy (ACC), Recall (REC), Specificity (SPE), Precision (PRE), F1-Score (F1).

| Models/Fold | | TP | TN | FP | FN | ACC (%) | REC (%) | SPE (%) | PRE (%) | F1 (%) |
|---|---|---|---|---|---|---|---|---|---|---|
| **InceptionV3** | Fold1 | 60 | 519 | 41 | 8 | 92.2 | 88.2 | 92.7 | 59.4 | 71.0 |
| | Fold2 | 60 | 547 | 13 | 8 | 96.7 | 88.2 | 97.7 | 82.2 | 85.1 |
| | Fold3 | 65 | 560 | 0 | 3 | 99.5 | 95.6 | 100 | 100 | 97.7 |
| | Fold4 | 57 | 524 | 36 | 11 | 92.5 | 83.8 | 93.6 | 61.3 | 70.8 |
| | Fold5 | 67 | 538 | 22 | 2 | 96.2 | 97.1 | 96.1 | 75.3 | 84.8 |
| | Total / Average | 309 | 2688 | 112 | 32 | 95.4 | 90.6 | 96.0 | 73.4 | 81.1 |
| **ResNet50** | Fold1 | 65 | 511 | 49 | 3 | 91.7 | 95.6 | 91.3 | 57.0 | 71.4 |
| | Fold2 | 59 | 545 | 15 | 9 | 96.2 | 86.8 | 97.3 | 79.7 | 83.1 |
| | Fold3 | 62 | 556 | 4 | 6 | 98.4 | 91.2 | 99.3 | 93.9 | 92.5 |
| | Fold4 | 59 | 543 | 17 | 9 | 95.9 | 86.8 | 97.0 | 77.6 | 81.9 |
| | Fold5 | 68 | 549 | 11 | 1 | 98.1 | 98.6 | 98.0 | 86.1 | 91.9 |
| | Total / Average | 313 | 2704 | 96 | 28 | 96.1 | 91.8 | 96.6 | 76.5 | 83.5 |
| **ResNet101** | Fold1 | 50 | 543 | 17 | 18 | 94.4 | 73.5 | 97.0 | 74.6 | 74.1 |
| | Fold2 | 49 | 559 | 1 | 19 | 96.8 | 72.1 | 99.8 | 98.0 | 83.1 |
| | Fold3 | 68 | 541 | 19 | 0 | 97.0 | 100 | 96.6 | 78.2 | 87.7 |
| | Fold4 | 33 | 554 | 6 | 35 | 93.5 | 48.5 | 98.9 | 84.6 | 61.7 |
| | Fold5 | 67 | 553 | 7 | 2 | 98.6 | 97.1 | 98.8 | 90.5 | 93.7 |
| | Total / Average | 267 | 2750 | 50 | 74 | 96.1 | 78.3 | 98.2 | 84.2 | 81.2 |
| **ResNet152** | Fold1 | 15 | 547 | 13 | 53 | 89.5 | 22.1 | 97.7 | 53.6 | 31.3 |
| | Fold2 | 55 | 556 | 4 | 13 | 97.3 | 80.9 | 99.3 | 93.2 | 86.6 |
| | Fold3 | 55 | 555 | 5 | 13 | 97.1 | 80.9 | 99.1 | 91.7 | 85.9 |
| | Fold4 | 34 | 539 | 21 | 34 | 91.2 | 50.0 | 96.3 | 61.8 | 55.3 |
| | Fold5 | 64 | 528 | 32 | 5 | 94.1 | 92.8 | 94.3 | 66.7 | 77.6 |
| | Total / Average | 223 | 2725 | 75 | 118 | 93.9 | 65.4 | 97.3 | 74.8 | 69.8 |
| **Inception-ResNetV2** | Fold1 | 59 | 538 | 22 | 9 | 95.1 | 86.8 | 96.1 | 72.8 | 79.2 |
| | Fold2 | 59 | 523 | 37 | 9 | 92.7 | 86.8 | 93.4 | 61.5 | 72.0 |
| | Fold3 | 38 | 553 | 7 | 10 | 97.2 | 79.2 | 98.8 | 84.4 | 81.7 |
| | Fold4 | 48 | 516 | 44 | 20 | 89.8 | 70.6 | 92.1 | 52.2 | 60.0 |
| | Fold5 | 64 | 542 | 18 | 5 | 96.3 | 92.8 | 96.8 | 78.0 | 84.8 |
| | Total / Average | 268 | 2672 | 128 | 53 | 94.2 | 83.5 | 95.4 | 67.7 | 74.8 |

Secondly, when the results obtained for the data in Binary Class-2 (COVID-19 / Viral pneumonia classes) are evaluated, the training performances of the models given in Figure 7 and Figure 8 are quite high. It can be said that the accuracy values and loss values of the ResNet50 and ResNet101 models perform better than the other models. Performance values obtained through test data are shown in Figure 9. Here, the models' results on the test data are generally more stable. There is no oscillation except when there is excessive oscillation only in the first 3 epochs of the ResNet50 model. Detailed performances of the models are given in Table 3. It is clearly seen that quite high values are reached for each fold value. While 99.4% was reached in the detection of COVID-19, it is seen that 99.5% was reached in the detection of viral pneumonia.

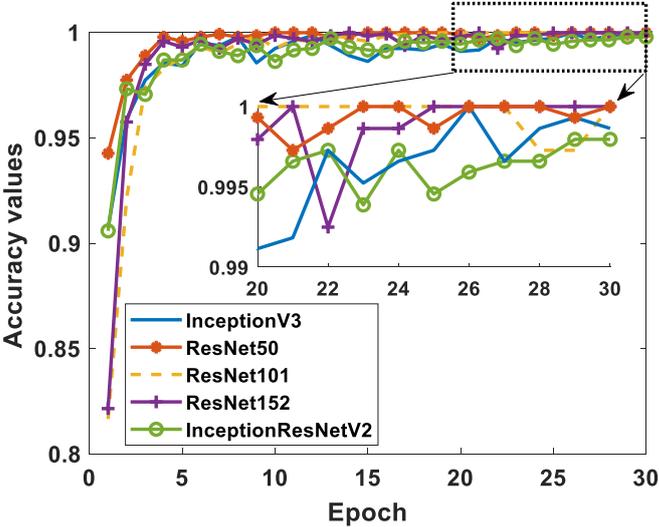

**Figure 7.** Binary Class-2: Comparison of training accuracy of 5 different models for fold-4.

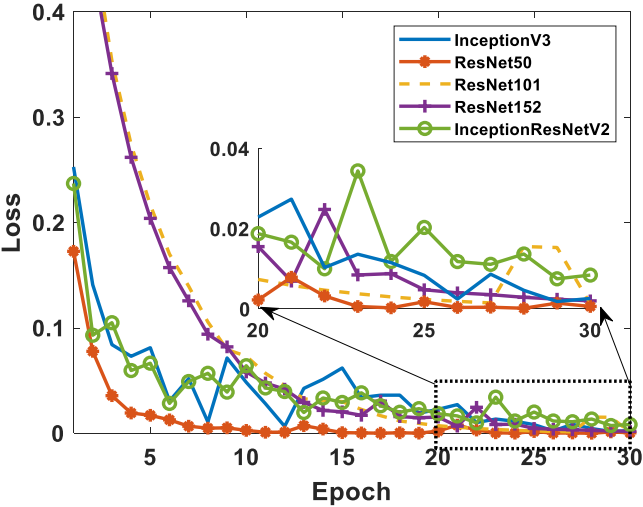

**Figure 8.** Binary Class-2: Comparison of training loss values of 5 different models for fold-4.

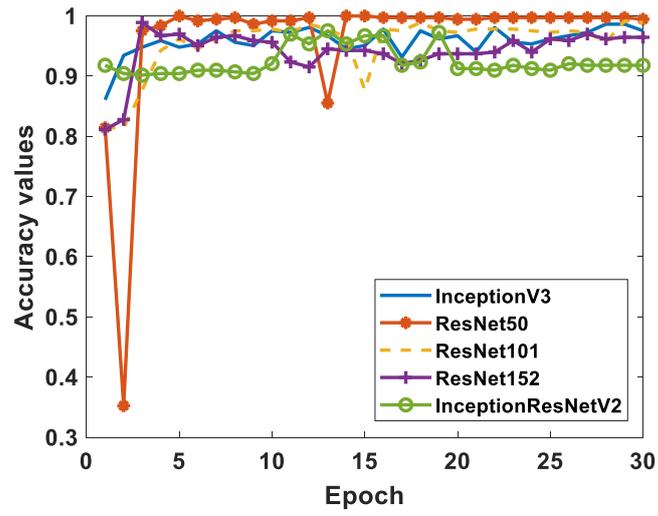

**Figure 9**. Binary Class-2: Comparison of testing accuracy of 5 different models for fold-4.

**Table 3**. All performances of 5 different models on each fold for COVID-19 / Viral pneumonia binary classification.

| Models/Fold | | TP | TN | FP | FN | ACC (%) | REC (%) | SPE (%) | PRE (%) | F1 (%) |
|---|---|---|---|---|---|---|---|---|---|---|
| **InceptionV3** | Fold1 | 68 | 292 | 6 | 0 | 98.4 | 100 | 98.0 | 91.9 | 95.8 |
| | Fold2 | 68 | 292 | 6 | 0 | 98.4 | 100 | 98.0 | 91.9 | 95.8 |
| | Fold3 | 67 | 295 | 4 | 1 | 98.6 | 98.5 | 98.7 | 94.4 | 96.4 |
| | Fold4 | 68 | 290 | 9 | 0 | 97.5 | 100 | 97.0 | 88.3 | 93.8 |
| | Fold5 | 69 | 299 | 0 | 0 | 100 | 100 | 100 | 100 | 100 |
| | Total/Average | 340 | 1468 | 25 | 1 | 98.6 | 99.7 | 98.3 | 93.2 | 96.3 |
| **ResNet50** | Fold1 | 68 | 297 | 1 | 0 | 99.7 | 100 | 99.7 | 98.6 | 99.3 |
| | Fold2 | 68 | 293 | 5 | 0 | 98.6 | 100 | 98.3 | 93.2 | 96.5 |
| | Fold3 | 68 | 298 | 1 | 0 | 99.7 | 100 | 99.7 | 98.6 | 99.3 |
| | Fold4 | 66 | 299 | 0 | 2 | 99.5 | 97.1 | 100 | 100 | 98.5 |
| | Fold5 | 69 | 299 | 0 | 0 | 100 | 100 | 100 | 100 | 100 |
| | Total/Average | 339 | 1486 | 7 | 2 | 99.5 | 99.4 | 99.5 | 98.0 | 98.7 |
| **ResNet101** | Fold1 | 61 | 294 | 4 | 7 | 97.0 | 89.7 | 98.7 | 93.8 | 91.7 |
| | Fold2 | 62 | 293 | 5 | 6 | 97.0 | 91.2 | 98.3 | 92.5 | 91.9 |
| | Fold3 | 68 | 295 | 4 | 0 | 98.9 | 100 | 98.7 | 94.4 | 97.1 |
| | Fold4 | 65 | 298 | 1 | 3 | 98.9 | 95.6 | 99.7 | 98.5 | 97.0 |
| | Fold5 | 45 | 299 | 0 | 24 | 93.5 | 65.2 | 100 | 100 | 78.9 |
| | Total/Average | 301 | 1479 | 14 | 40 | 97.1 | 88.3 | 99.1 | 95.6 | 91.8 |
| **ResNet152** | Fold1 | 63 | 291 | 7 | 5 | 96.7 | 92.6 | 97.7 | 90.0 | 91.3 |
| | Fold2 | 67 | 293 | 5 | 1 | 98.4 | 98.5 | 98.3 | 93.1 | 95.7 |
| | Fold3 | 66 | 298 | 1 | 2 | 99.2 | 97.1 | 99.7 | 98.5 | 97.8 |
| | Fold4 | 56 | 299 | 0 | 13 | 96.5 | 81.2 | 100 | 100 | 89.6 |
| | Fold5 | 58 | 298 | 1 | 10 | 97.0 | 85.3 | 99.7 | 98.3 | 91.3 |
| | Total/Average | 310 | 1479 | 14 | 31 | 97.5 | 90.9 | 99.1 | 95.7 | 93.2 |
| **Inception-ResNetV2** | Fold1 | 68 | 283 | 15 | 0 | 95.9 | 100 | 95.0 | 81.9 | 90.1 |
| | Fold2 | 68 | 267 | 31 | 0 | 91.5 | 100 | 89.6 | 68.7 | 81.4 |
| | Fold3 | 68 | 278 | 21 | 0 | 94.3 | 100 | 93.0 | 76.4 | 86.6 |
| | Fold4 | 41 | 296 | 3 | 27 | 91.8 | 60.3 | 99.0 | 93.2 | 73.2 |
| | Fold5 | 69 | 293 | 6 | 0 | 98.4 | 100 | 98.0 | 92.0 | 95.8 |
| | Total/Average | 314 | 1417 | 76 | 27 | 94.4 | 92.1 | 94.9 | 80.5 | 85.9 |

In the last study, the detection success of Binary Class-3 (COVID-19 / Bacterial pneumonia classes) was investigated. The performances of 5 different models on both training and test data are given in Figure 10, Figure 11 and Figure 12. As in other studies, it is clearly seen that the ResNet50 model exhibits higher training performance. InceptionV3 model is seen to exhibit increasing performance towards the end of the epoch number. When the detailed results given in Table 4 are evaluated, it can be said that the InceptionV3 model has a performance of 100% in the detection of COVID-19, while the overall performance is also said to be the ResNet50 model has a high success.

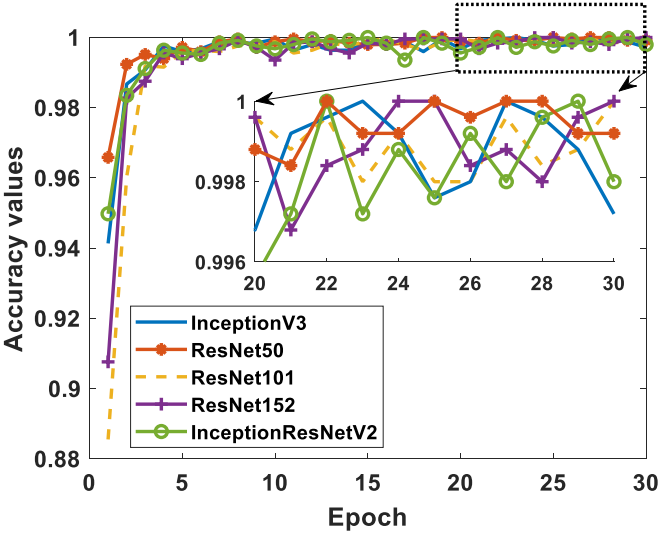

**Figure 10.** Binary Class-3: Comparison of training accuracy of 5 different models for fold-4.

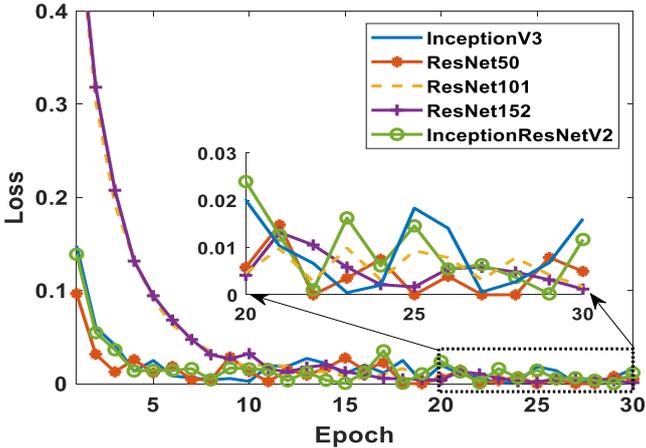

**Figure 11.** Binary Class-3: Comparison of training loss values of 5 different models for fold-4.

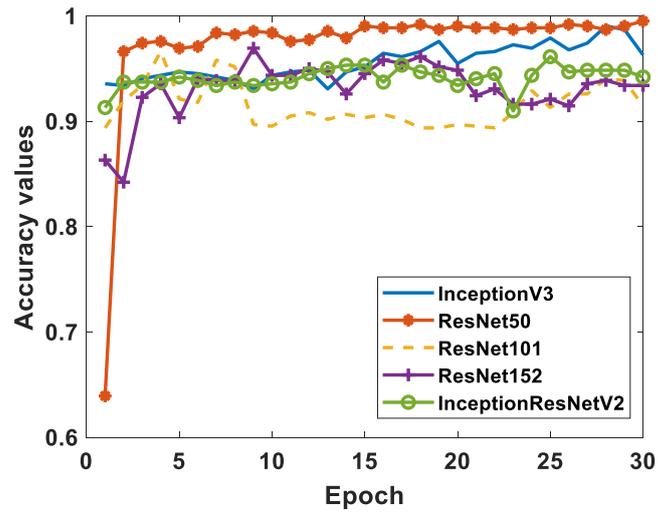

**Figure 12.** Binary Class-3: Comparison of testing accuracy of 5 different models for fold-4.

**Table 4.** All performances of 5 different models on each fold for COVID-19 / Bacterial pneumonia binary classification.

| Models/Fold | | TP | TN | FP | FN | ACC (%) | REC (%) | SPE (%) | PRE (%) | F1 (%) |
|---|---|---|---|---|---|---|---|---|---|---|
| **InceptionV3** | Fold1 | 68 | 554 | 0 | 0 | 100 | 100 | 100 | 100 | 100 |
| | Fold2 | 68 | 551 | 3 | 0 | 99.5 | 100 | 99.5 | 95.8 | 97.8 |
| | Fold3 | 68 | 541 | 13 | 0 | 97.9 | 100 | 97.7 | 84.0 | 91.3 |
| | Fold4 | 68 | 532 | 23 | 0 | 96.3 | 100 | 95.9 | 74.7 | 85.5 |
| | Fold5 | 69 | 521 | 34 | 0 | 94.6 | 100 | 93.9 | 67.0 | 80.2 |
| | Total/Average | 341 | 2699 | 73 | 0 | 97.7 | 100 | 97.4 | 82.4 | 90.3 |
| **ResNet50** | Fold1 | 68 | 554 | 0 | 0 | 100 | 100 | 100 | 100 | 100 |
| | Fold2 | 67 | 551 | 3 | 1 | 99.4 | 98.5 | 99.5 | 95.7 | 97.1 |
| | Fold3 | 68 | 554 | 0 | 0 | 100 | 100 | 100 | 100 | 100 |
| | Fold4 | 65 | 555 | 0 | 3 | 99.5 | 95.6 | 100 | 100 | 97.7 |
| | Fold5 | 69 | 552 | 3 | 0 | 99.5 | 100 | 99.5 | 95.8 | 97.9 |
| | Total/Average | 337 | 2766 | 6 | 4 | 99.7 | 98.8 | 99.8 | 98.3 | 98.5 |
| **ResNet101** | Fold1 | 42 | 554 | 0 | 26 | 95.8 | 61.8 | 100 | 100 | 76.4 |
| | Fold2 | 33 | 553 | 1 | 35 | 94.2 | 48.5 | 99.8 | 97.1 | 64.7 |
| | Fold3 | 68 | 554 | 0 | 0 | 100 | 100 | 100 | 100 | 100 |
| | Fold4 | 14 | 554 | 1 | 54 | 91.2 | 20.6 | 99.8 | 93.3 | 33.7 |
| | Fold5 | 22 | 555 | 0 | 47 | 92.5 | 31.9 | 100 | 100 | 48.4 |
| | Total/Average | 179 | 2770 | 2 | 162 | 94.7 | 52.5 | 99.9 | 98.9 | 68.6 |
| **ResNet152** | Fold1 | 9 | 554 | 0 | 59 | 90.5 | 13.2 | 100 | 100 | 23.4 |
| | Fold2 | 64 | 552 | 2 | 4 | 99.0 | 94.1 | 99.6 | 97.0 | 95.5 |
| | Fold3 | 26 | 554 | 0 | 42 | 93.2 | 38.2 | 100 | 100 | 55.3 |
| | Fold4 | 28 | 554 | 1 | 40 | 93.4 | 41.2 | 99.8 | 96.6 | 57.7 |
| | Fold5 | 47 | 502 | 53 | 22 | 88.0 | 68.1 | 90.5 | 47.0 | 55.6 |
| | Total/Average | 174 | 2716 | 56 | 167 | 92.8 | 51.0 | 98.0 | 75.7 | 60.9 |
| **Inception-ResNetV2** | Fold1 | 60 | 540 | 14 | 8 | 96.5 | 88.2 | 97.5 | 81.1 | 84.5 |
| | Fold2 | 39 | 552 | 2 | 29 | 95.0 | 57.4 | 99.6 | 95.1 | 71.6 |
| | Fold3 | 66 | 547 | 7 | 2 | 98.6 | 97.1 | 98.7 | 90.4 | 93.6 |
| | Fold4 | 34 | 551 | 4 | 34 | 93.9 | 50.0 | 99.3 | 89.5 | 64.2 |
| | Fold5 | 42 | 536 | 19 | 27 | 92.6 | 60.9 | 96.6 | 68.9 | 64.6 |
| | Total/Average | 241 | 2726 | 46 | 100 | 95.3 | 70.7 | 98.3 | 84.0 | 76.8 |

## 5. Discussion

The use of artificial intelligence-based systems is very common in detecting those caught in the COVID-19 epidemic. As given in Table 5, there are many studies on this subject in the literature. In binary classification, it is common to distinguish COVID-19 positive from COVID-19 negative. In addition, it is very important to distinguish viral and bacterial pneumonia patients, which are other types of diseases affecting the lung, from COVID-19 positive patients. There are a limited number of studies in the literature that work with multiple classes. Das et al. conducted studies for 3 different classes (COVID-19 positive, pneumonia, and other infection). The researchers used 70% of the data for the training, the remaining 10% for validation and 20% for the test. As a result, they obtained 94.40% accuracy over test data with the CNN model they suggested [9]. Singh et al. proposed a two-class study using limited data. They reported their performances by dividing the dataset at different training and testing rates. They achieved the highest accuracy of 94.65 ± 2.1 at 70% training - 30% testing rates. In their study, they set the CNN hyper-parameters using multi-objective adaptive differential evolution (MADE) [49]. Afshar et al. conducted their studies using a method called COVID-CAPS with multi-class (Normal, bacterial pneumonia, Non-COVID viral pneumonia and COVID-19) studies. They achieved 95.7% accuracy with the approach without pre-training and 98.3% accuracy with pre-trained COVID-CAPS. However, although their sensitivity values are not as high as general accuracy, they detected the without pre-training and 98.3% accuracy with pre-trained COVID-CAPS as 90% and 80%, respectively [31].

Ucar and Korkmaz carried out multi-class (Normal, pneumonia and COVID-19 cases) work with deep Bayes-SqueezeNet. They obtained the average accuracy value of 98.26%. They worked with 76 COVID-19 data [25]. Sahinbas and Catak worked with 5 different pre-trained models (VGG16, VGG19, ResNet, DenseNet, and InceptionV3). They achieved 80% accuracy with VGG16 as their binary classifier performances. They worked with 70 COVID-positive and 70 COVID-negatives in total [27]. Khan et al. worked with normal, pneumonia-bacterial, pneumonia-viral and COVID-19 chest X-ray images. As a result, they achieved 89.6% overall performance with the model they named CoroNet. They used 290 COVID-19 data. They worked with more COVID-19 data than many studies [22]. Medhi et al. achieved 93% overall performance value in their study using deep CNN. They worked with 150 pieces of COVID-19 data [28].

In another study, Zhang and his colleagues performed binary and multi-class classifications containing 106 COVID-19 data. They found the detection accuracy of 95.18% with the confidence-aware anomaly detection (CAAD) model [17]. Apostopolus et al. obtained an accuracy of 93.48% using a total of 224 COVID-19 data with the VGG-19 CNN model for their 3-classes (COVID-19 -bacterial-normal) study [26]. Narin et al. used 50 COVID-19 / 50 Normal data in their study, where they achieved 98% accuracy with ResNet50 [32]. In many studies in the literature, researchers have studied a limited number of COVID-19 data. In this study, the differentiation performance of 341 COVID-19 data from each other was investigated with 3 different studies. In the study, 5 different CNN models were compared. The most important points in the study can be expressed as follows:

- There is no manual feature extraction, feature selection and classification in this method. It was realized end-to-end directly with raw data.
- The performances of the COVID-19 data across normal, viral pneumonia and bacterial pneumonia classes were significantly higher.
- It has been studied with more data than many studies in the literature.
- It has been studied and compared with 5 different CNN models.
- A high-accuracy decision support system has been proposed to radiologists for the automatic diagnosis and detection of patients with suspected COVID-19 and follow-up.

From another point of view, considering that this pandemic period affects the whole world, there is a serious increase in the work density of radiologists. In these manual diagnoses and determinations, the expert's tiredness may increase the error rate. It is clear that decision support systems will be needed in order to eliminate this problem. Thus, a more effective diagnosis can be made.

The most important issue that restricts this study is to work with limited data. Increasing the data, testing it with the data in many different centers will enable the creation of more stable systems.

In future studies, the features will be extracted using image processing methods on X-ray and CT images. From these extracted features, the features that provide the best separation between classes will be determined and performance values will be measured with different classification algorithms. In addition, the results will be compared with deep learning models. Apart from this, the results of the study will be tested with data from many different centers. In a future study, studies will be conducted to determine the demographic characteristics of patients and COVID-19 possibilities with artificial intelligence-based systems.

**Table 5.** The performance comparison literature about COVID-19 diagnostic methods using chest X-ray images.

| Previous Study | Data Type | Methods / Classifier | Number of Classes | Accuracy (%) |
|---|---|---|---|---|
| Das et al. [9] | X-ray | Xception | 3 | 97.40 |
| Singh et al. [49] | X-ray | MADE based CNN | 2 | 92.55 |
| Afshar et al. [31] | X-ray | Capsule Networks | 4 | 95.7 |
| Ucar and Korkmaz [25] | X-ray | Bayes-SqueezeNet | 3 | 98.3 |
| Khan et al. [22] | X-ray | CoroNet | 4 | 89.60 |
| Sahinbas and Catak [27] | X-ray | VGG16, VGG19, ResNet, DenseNet, InceptionV3 | 2 | 80 |
| Medhi et al. [28] | X-ray | Deep CNN | 2 | 93 |
| Zhang et al. [17] | X-ray | CAAD | 2 | 95.18 |
| Apostopolus et. al. [26] | X-ray | VGG-19 | 3 | 93.48 |
| Narin et al. [32] | X-ray | **InceptionV3, ResNet50, Inception-ResNetV2** | 2 | 98 |
| **This Study** | **X-ray** | **InceptionV3, ResNet50, ResNet101, ResNet152, Inception-ResNetV2** | **2 (COVID-19 / Normal)** | **96.1** |
| **This Study** | **X-ray** | **InceptionV3, ResNet50, ResNet101, ResNet152, Inception-ResNetV2** | **2 (COVID-19 / Viral Pneumonia)** | **99.5** |
| **This Study** | **X-ray** | **InceptionV3, ResNet50, ResNet101, ResNet152, Inception-ResNetV2** | **2 (COVID-19 / Bacterial Pneumonia)** | **99.7** |

## 6. Conclusion

Early prediction of COVID-19 patients is vital to prevent the spread of the disease to other people. In this study, we proposed a deep transfer learning based approach using Chest X-ray images obtained from normal, COVID-19, bacterial and viral pneumonia patients to predict COVID-19 patients automatically. Performance results show that ResNet50 pre-trained model yielded the highest accuracy among five models for used three different datasets (Dataset-1: 96.1%, Dataset-2: 99.5% and Dataset-3: 99.7%) . In the light of our findings, it is believed that it will help radiologists to make decisions in clinical practice due to the higher performance. In order to detect COVID-19 at an early stage, this study gives insight on how deep transfer learning methods can be used. In subsequent studies, the classification performance of different CNN models can be tested by increasing the number of COVID-19 Chest X-ray images in the dataset.